\documentclass[preprint2,lineno]{aastex6}

\usepackage{amsmath}	
\usepackage{xspace}
\usepackage{color}
\usepackage{multirow}
\usepackage{ifpdf}
\usepackage[flushleft]{threeparttable}

\newcommand{\gray}{$\gamma$-ray\xspace}
\newcommand{\grays}{$\gamma$-rays\xspace}
\newcommand{\unit}[1]{\,\mathrm{#1}\xspace}
\newcommand{\Fermi}{\emph{Fermi}\xspace}
\newcommand{\FermiLAT}{\emph{Fermi}~LAT\xspace}
\newcommand{\fermiLAT}{\emph{Fermi}-LAT\xspace}
\newcommand{\elmag}{\texttt{ELMAG}\xspace}

\begin{document}
\title[]{Sensitivity of the Cherenkov Telescope Array to the 
Detection of Intergalactic Magnetic Fields}

\author{Manuel Meyer and Jan Conrad}
\affil{The Oskar Klein Center for CosmoParticle Physics, Department of Physics, Stockholm University, Albanova, SE-10691 Stockholm, Sweden} \email{manuel.meyer@fysik.su.se}
\email{conrad@fysik.su.se}
\and
\author{Hugh Dickinson}
\affil{Department of Physics \& Astronomy, Iowa State University, Ames, IA 50011-3160, USA}
\email{hughd@iastate.edu}

\begin{abstract}
Very high energy (VHE; energy $E \gtrsim 100\,$GeV) \grays originating from extragalactic sources 
undergo pair production with low-energy photons of background radiation fields.
These pairs can inverse-Compton-scatter background photons, 
initiating an electromagnetic cascade.
The spatial and temporal structure of this secondary \gray signal is altered as the   $e^+e^-$ pairs are deflected in 
an intergalactic magnetic field (IGMF).
We investigate how VHE observations with the future Cherenkov Telescope Array,  with its high angular resolution and broad energy range, can potentially probe the IGMF.
We identify promising sources and simulate \gray spectra
 over a wide range of values of the IGMF strength
and coherence length using the publicly available {\elmag} Monte Carlo code.  
Combining simulated observations in a joint likelihood approach,
we find that current limits 
on the IGMF can be significantly improved.
The projected sensitivity depends strongly on the time a source has been \gray active and on the emitted maximum \gray energy.
\end{abstract}


\section{Introduction}
The origin of the magnetic fields ubiquitously present in galaxies, galaxy clusters, 
and perhaps filaments of large-scale structure 
is yet unknown. 
The general consensus is that
observed fields can (at least partially)
be explained by
pre-existing fields
that are
amplified during the gravitational collapse of forming structures
via flux compression and dynamos \citep[see, e.g.,][for reviews]{widrow2002,kulsrud2008,durrer2013}.
However, little is known about the required seed fields.
On the one hand, they could have formed in the very early Universe during the electroweak or QCD phase transition \citep[e.g.,][]{grasso2001,widrow2002} 
or during inflation \citep[e.g.,][]{durrer2013}.
On the other hand, the seed fields could be of astrophysical origin and could have been produced
during the formation of large-scale structures at redshifts $z \lesssim 10$.
The voids could have been polluted by magnetic fields through galactic outflows caused by star formation \citep{bertone2006} 
or active galactic nuclei \citep[AGNs;][]{rees1987,daly1990,ensslin1997,furlanetto2001}.
The two scenarios could be discerned by measuring the strength of intergalactic magnetic fields (IGMFs) in voids,
 $B$, and their coherence length $\lambda$ simultaneously.
The coherence lengths of primordial fields should be $\lambda \lesssim\mathcal{\mathrm{kpc}}$, whereas
astrophysical fields should have field strengths $B \lesssim10^{-9}\,\mathrm{G}$ 
with $\lambda\gtrsim\mathcal{\mathrm{kpc}}$ \citep{durrer2013}.

So far, no direct measurements  of $B$ and $\lambda$ exist. 
The nonobservation of Faraday rotation induced by an IGMF in quasar observations
leads to upper limits of $B\lesssim10^{-9}\,\mathrm{G}$ for megaparsec-scale coherence lengths \citep{blasi1999,pshirkov2015}.
If the IGMF is of primordial origin, it will affect the primordial plasma and will leave specific imprints on the spectrum
and polarization of the cosmic microwave background (CMB). 
\textit{Planck} observations yield limits on the primordial IGMF
with $B\lesssim10^{-9}\,\mathrm{G}$ for $\lambda =1\unit{Mpc}$,
where the exact value depends on the considered IGMF model \citep{planck2015bfields}.
Simulations of the formation of galaxy clusters and the propagation of 
ultrahigh energy cosmic rays suggest lower values of the order of $B \lesssim 10^{-12}\,\mathrm{G}$ \citep[][]{sigl2004,dolag2005}.

Observation of \grays originating from blazars, 
AGNs with their jet closely aligned along the line of sight to the observer,
provide an independent probe of the IGMF.
The very high energy (VHE; energy $E \gtrsim 100\,\mathrm{GeV}$) flux of AGN is attenuated due to the interaction of \grays
with photons of the extragalactic background light (EBL), 
$\gamma + \gamma_\mathrm{EBL} \to e^+ + e^-$ \citep[]{nikishov1962,jelley1966,gould1967,gould1967a,dwek2013}.
The attenuation scales exponentially with the optical depth
$\tau(E,z)$, a monotonically increasing function with 
both the primary \gray energy $E$ and the source redshift $z$.
The produced pairs can inverse-Compton (IC) scatter photons of the CMB and EBL
 and induce an electromagnetic cascade \citep{protheroe1993}.
As the pairs are deflected in the magnetic field, the angular and time structure of the secondary 
photon signal depends on the strength and morphology of the IGMF.
Under the assumption of a certain EBL model and intrinsic source spectrum, the nonobservation of the cascade component at GeV energies 
with the \textit{Fermi} Large Area Telescope (LAT) led to a lower limit of $B \gtrsim 10^{-16}\,\mathrm{G}$ for 
$\lambda = 1\unit{Mpc}$ \citep{neronov2010,tavecchio2011},
or conversely a lower limit on the the filling factor of the IGMF along the line of sight \citep{dolag2011}.
Additionally, the IGMF induces a time delay of the cascade emission 
compared to the primary source emission \citep{plaga1995,dai2002,murase2008,takahashi2008,neronov2009}. 
If this is taken into account, together with conservative assumptions on the AGN \gray activity, 
the limit is relaxed by several orders of magnitude, 
$B \gtrsim 10^{-19}\,\mathrm{G}$, 
as derived from semianalytical models \citep{dermer2011,huan2011,finke2015}
and full Monte Carlo simulations for simultaneous observations with imaging air Cherenkov telescopes 
(IACTs) and the \textit{Fermi} LAT,
leading to $B \gtrsim 10^{-17}\,\mathrm{G}$ \citep{taylor2011}.
For small EBL photon densities and taking uncertainties of the intrinsic source spectrum into account, 
the hypothesis of a zero IGMF cannot be rejected \citep{arlen2014}. 

Strong magnetic fields could also be detected through the angular profile of the \gray
 emission since the $e^+e^-$ 
pairs would be quickly isotropized and extended \gray halos 
would form around sources \citep{aharonian1994,dolag2009,elyiv2009,neronov2009}.  
No extended emission was found in H.E.S.S. observations \citep{hess2014ph},
whereas indications for pair halos were found in \fermiLAT data,
suggesting magnetic fields of the order of $10^{-17}\unit{G}\lesssim B \lesssim10^{-15}\unit{G}$
for $\lambda = 1\unit{Mpc}$ \citep{chen2015ph}. 

A helical IGMF could be detected through parity odd signatures of the arrival directions
of \grays produced in the electromagnetic cascade \citep{tashiro2013}.
An analysis of the diffuse \gray background observed with 
\FermiLAT indeed suggests such correlations with favored magnetic fields 
$B \sim 5\times10^{-14}\unit{G}$ ordered over 10\,Mpc scales \citep{chen2015helical}.

The cascade could be suppressed if the $e^+e^-$ pairs could lose their 
energy primarily via plasma instabilities instead of IC scattering \citep{broderick2012}.
Particle-in-cell simulations suggest that the energy loss
due to the instabilities plays only a subdominant role \citep{sironi2014}.
However, as noted by \citet{menzler2015}, 
these simulations require extrapolations 
over many orders of magnitude in the density ratio between the
beam and the background plasma.
The authors find instead a reduction of the cascade flux by a factor of 0.1
for typical blazars.

Assuming no energy losses in plasma instabilities,
 we investigate the prospects of the future Cherenkov Telescope Array (CTA) to detect the secondary cascade emission.
The amount of cascade emission that arrives within a certain 
maximum delay time and within the CTA point spread function (PSF)
depends on the IGMF, thereby allowing constraints on its 
strength \citep[e.g.][]{taylor2011}. We follow a similar 
approach here.

CTA will be composed of IACTs of different sizes, covering a large energy range
between tens of GeV up to hundreds of TeV 
with an expected sensitivity improvement of a factor of 10
 compared to currently 
operating IACTs \citep{cta2011}.
The energy resolution is envisaged to be of the order of 
10\,\%-15\,\% and the spatial resolution of the order of arcminutes.  
The broad energy coverage makes it possible to detect the primary and secondary spectral components simultaneously. 
Throughout this paper, the ``Array E'' configuration of CTA will be assumed,
 which provides 
good sensitivity over the entire energy range \citep{bernloehr2013}.
Current limits often rely on the combination of IACT and \fermiLAT data,
which are usually nonsimultaneous and suffer from the systematic uncertainty of potentially different energy scales. 
These issues are avoided with CTA observations. 

The article is organized as follows. In Section \ref{sec:sources} we present our 
source selection for promising blazars to search for the cascade. 
As described in Sec.
tion \ref{sec:sim}, we use the publicly available 
\elmag Monte Carlo code to generate spectra including a cascade for a wide range 
of IGMF morphologies. 
We use a standard likelihood ratio test to search for the cascade component (Section \ref{sec:analysis})
and combine observations in a joint likelihood.
We present our results in Section \ref{sec:results} before concluding in Section \ref{sec:conclusion}.

\section{Source Selection}
\label{sec:sources}
Promising targets to search for the cascade are blazars whose 
intrinsic emission extends to energies where the EBL absorption is strong. 
This requires sources with a hard intrinsic spectrum characterized 
by a simple power law without a cutoff, $\mathrm{d}N/\mathrm{d} E \propto E^{-\Gamma}$,
with $\Gamma \lesssim 2$. 
In addition to being a high synchrotron
 peaked BL\,Lac (HBL,  $\log_{10}(\nu_\mathrm{sync} / \mathrm{Hz}) > 15$),
promising sources also show a high ratio between the X-ray and radio flux, $F_\mathrm{X} / F_\mathrm{R} \gtrsim 10^4$,
 as well as an optical spectrum dominated by the host galaxy
 \citep{bonnoli2015}.
\citeauthor{bonnoli2015} used values for 
the X-ray and radio flux as listed in \citet{plotkin2010}, 
with $F_\mathrm{X}$
measured with \emph{ROSAT} between $0.1$ and $2.4\,$keV 
and $F_\mathrm{R} = \nu F_\nu$ with $\nu = 1.4\,$GHz.
1ES\,0229+200 is a typical example  for such an ``extreme'' HBL (EHBL; \citealt{costamante2001}).

We follow these requirements and 
select sources from the second \Fermi catalog of hard sources detected above 50\,GeV
\citep[2FHL;][]{2fhl} that fulfill the following criteria:

\begin{enumerate}
\setlength\itemsep{-0.9pt}
\item Their redshift is known. Otherwise, it is not possible to determine the strength of the
absorption.
 \item They are HBLs, i.e.  $\log_{10}(\nu_\mathrm{sync} / \mathrm{Hz}) > 15$.
 \item They show a high ratio between their X-ray and radio flux, $F_\mathrm{X} / F_\mathrm{R} > 10^3$, 
 where the fluxes are taken from the third \fermiLAT AGN catalog \citep[3LAC;][]{3lac}. 
 The X-ray flux is taken from the \emph{ROSAT} all sky survey  between $0.1$ and $2.4$\,keV
 and the the radio flux is determined from $F_\mathrm{R} = \nu F_\nu$. 
 The frequency varies depending on the radio survey used
 (see Table 8 in the 3LAC).  
\item Integrating their observed 2FHL spectra between 1 and 2\,TeV 
should result in at least 1\,\% of the integrated flux of the Crab Nebula in the same energy range
(assuming the VHE Crab spectrum measured with H.E.S.S.; \citealt{crabhess2006}). 
\item The absorption-corrected spectra in the 2FHL \citep{dominguez2015} follow  power laws with an index $\Gamma \leqslant 1.7$.
This value is chosen \textit{a posteriori}, as softer spectra do not turn out to lead to a sizable flux of the cascade photons. 
By making this cut, we assume that the \fermiLAT observations are not contaminated by the cascade. 
As we will see in Sec. \ref{sec:sim}, this is justified by taking the 2FHL error bars of most 
sources into account.
\item  The sources should show little \gray variability as we assume a steady \gray emission to calculate the cascade. Following \citet{finke2015}, we select sources with  a variability index $< 100$ as provided in the third \Fermi source catalog \citep[3FGL;][]{3fgl}. This corresponds to a 4.8$\,\sigma$ significance that the source is variable.
\item They culminate at low zenith angles, $Z \leqslant 20^\circ$, in order to guarantee 
an energy threshold as low as possible. For this selection, we assume latitudes of $-24^\circ$ and $29^\circ$
 for the southern and northern CTA site, respectively.  
\end{enumerate}

In addition to the above criteria, we demand that the cascade photons have an energy $> 50\,$GeV. 
As the IC scattering with CMB photons with an average energy $\langle \epsilon_\mathrm{CMB} \rangle \approx 634\,\mu$eV
occurs entirely in the Thomson regime, 
the average energy of a cascade photon is
\begin{equation}
 \langle \epsilon \rangle = \frac{4}{3} \langle \epsilon_\mathrm{CMB} \rangle \gamma^2\approx 0.81\left(\frac{E}{\mathrm{TeV}}\right)^2\,\mathrm{GeV},
 \label{eq:esecond}
\end{equation}
with an electron Lorentz factor $\gamma = E / 2m_ec^2$, 
where $m_e$ is the mass of the electron and $E$ the energy of the primary \gray.
We estimate the maximum cascade photon energy from head-on IC scattering 
in the Thompson regime with a CMB photon with an energy of $\epsilon_{\mathrm{CMB},99} \approx 2$\,meV. 
The integral over the CMB photon density up to this energy is equal to 99\,\%
of the same integral between $[0;\infty)$.
One finds a maximum cascade energy
 \begin{equation}
\epsilon_\mathrm{max} = 4 \epsilon_{\mathrm{CMB},99} \gamma^2\approx 9 \,\langle \epsilon \rangle.
 \label{eq:emax}
\end{equation}
To decide whether $\epsilon_\mathrm{max}$ falls inside the CTA energy range, one 
has to make an assumption about the maximum energy of the primary \gray spectrum. 
A primary \gray spectrum that extends to high energies will also lead 
to more energy that can be reprocessed in the cascade.
Evidence for emission at energies beyond $\tau > 5$ has been found in several blazar observations,
e.g. for 1ES\,0229+200 \citep{1es0229hess2007}, PKS\,1424+240 \citep{pks1424veritas2014}
with $z \geqslant 0.6035$ \citep{furniss2013},
 and PKS\,0447--121 \citep{pks0447hess2013} 
assuming the redshift of $z = 0.343\pm0.002$  \citep{muriel2015}.
We therefore assume that the spectrum extends to an energy where the optical depth $\tau = 5$
(we will scrutinize this assumption in Sec. \ref{sec:results}).

In total, nine HBLs listed in the 2FHL survive the applied cuts. From this list we further exclude 
 the already TeV-detected sources IC\,310, RBS\,0413, RX\,J0648.7+1516, 1RXS\,J101015.9--311909, and B3\,2247+381.
None of the IACT spectra extend to high optical depths, and the measured indices 
are significantly softer than the ones listed in the 2FHL. 
The remaining four blazars are listed in Table \ref{tab:srcs}. We append 1ES\,0229+200 
to the list, even though the source is not included in the 2FHL. 

\begin{table*}
\centering
\caption{\label{tab:srcs}Sources selected for Simulation.}
\begin{footnotesize}
\begin{tabular}{lccccccccc}
\hline
\hline
\multirow{2}{*}{Source name} & R.A. & Decl. & \multirow{2}{*}{$z$} 
& $E_\mathrm{HEP}$ & $E_{\tau = 5}$ & \multirow{2}{*}{$\log_{10}\left(\frac{\nu_\mathrm{sync}}{\mathrm{Hz}}\right)$} &
$F\,(10^{-10}\,\mathrm{ergs}\,\mathrm{cm}^{-2}\,\mathrm{s}^{-1})$
& \multirow{2}{*}{$\Gamma \pm \sigma_\Gamma$} 
& \multirow{2}{*}{$\frac{F_\mathrm{X}}{F_\mathrm{R}}$} \\ 
{} & (deg) & (deg) & {} & (GeV) & (TeV) & {} & (Above $E_\mathrm{thr} / \mathrm{TeV}$) & {} & {} \\
\hline
B2\,0806+35               &$    122.39$&$     34.97$&$     0.083$&$    264.23$&$    13.931$&$    15.500$&$    7.24 \,(1.0)$&$  0.920\pm  0.770$&$   3158$\\
PG\,1218+304              &$    185.34$&$     30.16$&$     0.182$&$    513.20$&$     6.488$&$    16.590$&$    2.43 \,(0.2)$&$  1.630\pm  0.270$&$  31508$\\
PMN\,J1548--2251           &$    237.19$&$    -22.82$&$     0.192$&$    435.85$&$     5.935$&$    16.061$&$     1.02 \,(0.5)$&$  1.340\pm  0.490$&$  15455$\\
1RXS\,J023832.6--311658    &$     39.62$&$    -31.27$&$     0.232$&$    407.77$&$     4.091$&$    16.160$&$    0.57 \,(0.5)$&$  0.760\pm  0.650$&$   9050$\\
\hline
1ES\,0229+200           &$    38.20$&$	20.29$&$     0.139$&-&$      9.242$&$    15.481$&$1.35 \,(0.58)$&$1.7$&$  13430$\\
\hline
\end{tabular}
\begin{tablenotes}
\item \textbf{Note.}
 In addition to the source coordinates we give the redshift $z$
(the redshift of PMN\,J1548--2251 is taken from \citealt{shaw2013}),
the highest energy photon, $E_\mathrm{HEP}$, the energy at which 
$\tau = 5$, $E_{\tau = 5}$, the peak frequency of the synchrotron 
emission, the integrated energy flux $F$ between $E_\mathrm{thr}$ and $E_{\tau = 5}$ 
assumed for the simulation,
the spectral index of the intrinsic blazar spectrum $\Gamma$ with its uncertainty,
and the X-ray to radio flux ratio.
X-ray and radio fluxes are taken from the 3LAC. The values for the coordinates, $E_\mathrm{HEP}$, $\nu_\mathrm{sync}$, 
and $\Gamma$ (derived by de-absorbing the observed spectra with the EBL model of \citealt{dominguez2011}) 
are taken from the 2FHL. 
\end{tablenotes}
\end{footnotesize}
\end{table*}

\section{Simulations}
\label{sec:sim}

\subsection{Cascade Simulations}
The development of electromagnetic cascades in the intergalactic medium
is simulated with the open-source Monte Carlo code \elmag 
\citep[for more details see][]{elmag}. 
\elmag computes the resultant photon distribution based
 on initial prescriptions for the shape of the intrinsic \gray spectrum, 
the spectrum and redshift evolution of the EBL, and the large-scale configuration of the IGMF. 
It adopts the simplifying assumption that the IGMF can be completely characterized by a universal intensity $B$
 and a cell-like structure with coherence length $\lambda$.
The scattering of the $e^+e^-$ pairs on both EBL and CMB photons is taken into account. 
Photons with an energy 
above $\epsilon \geqslant \epsilon_\mathrm{thr}$ are traced, where we set $\epsilon_\mathrm{thr} = 1\unit{GeV}$.
Energy losses due to synchrotron radiation and IC scattering
 are integrated out for energies $<\epsilon_\mathrm{thr}$.   
 We choose to trace all particles, i.e. setting $\alpha_\mathrm{sample} = 0$ \citep{elmag}.

Under the assumption that the $e^+e^-$ pairs do 
not lose energy in plasma instabilities, we simulate the final photon distribution from $6\times10^5$ injected primary 
\grays with the following fiducial model assumptions
 for each considered source:
\begin{enumerate}
\setlength\itemsep{-0.9pt}
\item The intrinsic \gray spectrum is given by a power law, $\mathrm{d} N / \mathrm{d} E\propto E^{-\Gamma}$, with the absorption-corrected index of the 2FHL (see Table \ref{tab:srcs}). For 1ES\,0229+200, 
we assume an intrinsic index $\Gamma = 1.7$. 
\item We choose the EBL model of \citet{dominguez2011}.
\item We assume that the primary \gray emission extends to an energy for which $\tau _\mathrm{max} = 5$. Beyond 
this energy, the emission is zero. 
\item An opening angle of the blazar jet of $\theta_\mathrm{j} = 6^\circ$ is assumed.
In the approximation that the Doppler factor is equal to the bulk Lorentz factor  $\Gamma_\mathrm{L}$ of the emitting plasma,
this implies $\Gamma_\mathrm{L} \sim \theta_\mathrm{j}^{-1} \sim 10$.
\item We simulate cascades for values of $(B,\lambda)$ on a $(9\times9)$ logarithmic spaced grid with 
$B_\mathrm{G} = B / \mathrm{G} \in [10^{-19}; 10^{-11}]$ and $\lambda_\mathrm{Mpc} = \lambda / \mathrm{Mpc}
\in [10^{-6}; 10^2]$. These parameters cover the evidences for a nonzero IGMF \citep{chen2015ph,chen2015helical},
as well as scenarios for astrophysical
or primordial origins of the IGMF \citep[e.g.,][]{durrer2013}.
\end{enumerate}
We discuss the impact of these assumptions in Sec. \ref{sec:results}.

\begin{figure}
\centering
\ifpdf
\includegraphics[width = .99\linewidth]{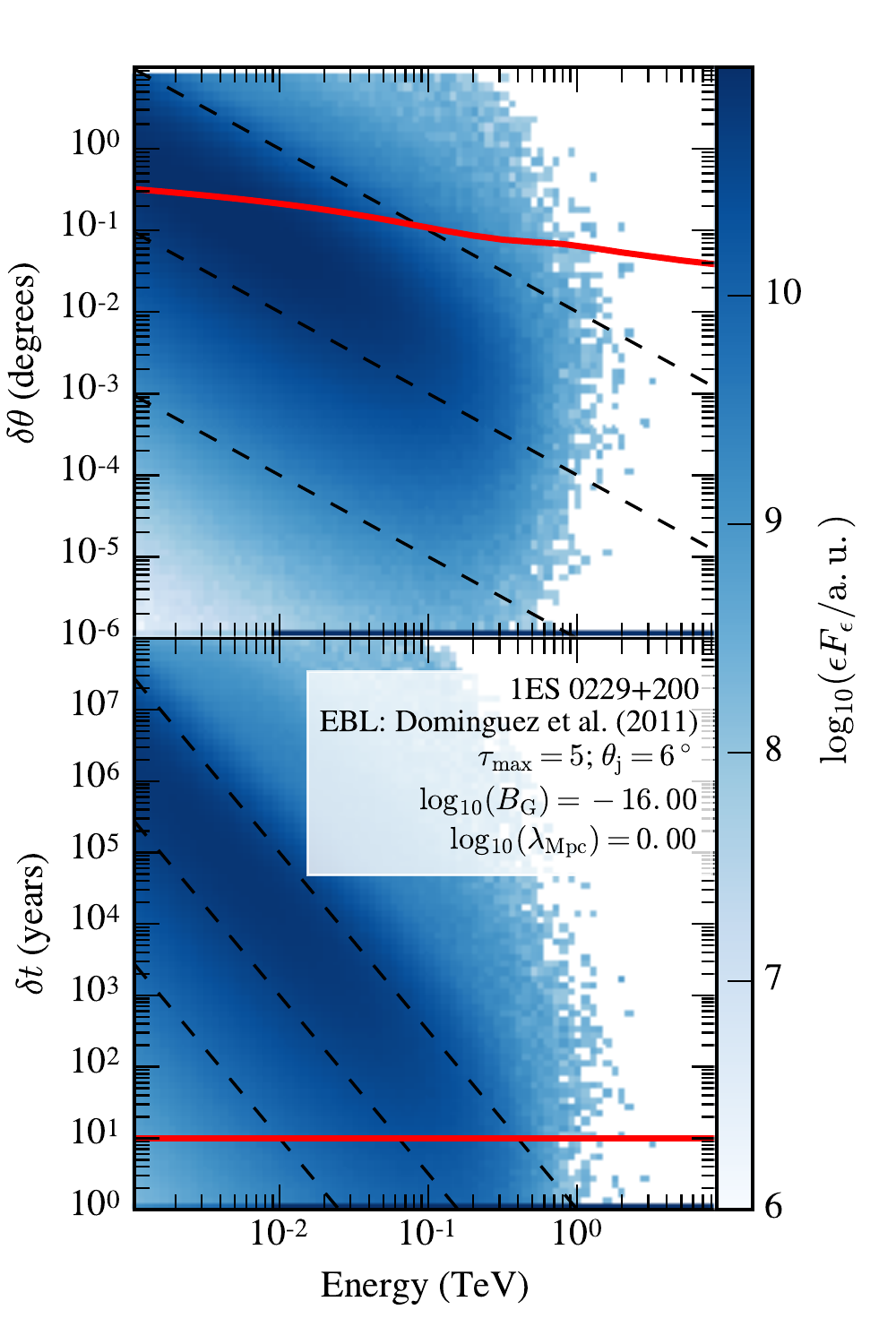}
\else
\vspace{10pt}
\includegraphics[width = .99\linewidth]{fig1.eps}
\fi
\caption{\label{fig:2dhist}
Two-dimensional histograms showing the output
of the \elmag Monte Carlo simulation. 
For the simulation, we assume the spectrum 
of  1ES\,0229+200, 
$B_\mathrm{G} = 10^{-16}$, and $\lambda_\mathrm{Mpc} = 1$.
All other parameters are set to their fiducial values. 
The \elmag output is binned in energy and $\delta\theta$ (top)
or $\delta t$ (bottom) and each bin shows the contained 
$\epsilon F_\epsilon$ in arbitrary units (a.u.). 
To obtain the energy spectra (see Figure \ref{fig:single_sources})
within the PSF containment radius (or maximum time delay), 
one has to sum the histogram along the $\delta \theta$ ($\delta t$)
axis for all entries below the red solid lines. 
Theoretical expectations for the energy dependence of 
$\delta\theta$ and $\delta t$ are shown as black dashed lines to guide the eye. 
}
\end{figure}

\elmag outputs the total observed spectrum $\epsilon F_\epsilon$ with primary and cascade emission
binned in energy, angular separation $\delta\theta$, and time delay $\delta t$. 
Both $\delta\theta$ and $\delta t$ are due to the deflection 
of the $e^+e^-$ pairs in the IGMF. 
We show an example of the output photon distribution as a function of energy 
and $\delta \theta$ ($\delta t$) for one pair of $(B,\lambda)$
values for 1ES\,0229+200 in Figure \ref{fig:2dhist}.
The distributions follow the theoretical expectations 
that $\delta\theta \propto \epsilon^{-1}B$ and $\delta t \propto
\epsilon^{-5/2}B^2$ since the chosen value of $\lambda = 1\unit{Mpc} \gtrsim D_\mathrm{IC}$, 
where $ D_\mathrm{IC}$ is the IC cooling length \citep{neronov2009}.
With the Thomson cross section $\sigma_T$ and the energy density 
of the CMB $u_\mathrm{CMB} = 0.26\unit{eV}\unit{cm}^{-3}$,
 the cooling length is 
\begin{equation}
 D_\mathrm{IC} = \frac{3m_ec^2}{4\sigma_\mathrm{T}u_\mathrm{CMB}\gamma} \approx 0.7 \left(\frac{E}{\mathrm{TeV}}\right)^{-1}\,\mathrm{Mpc}.
 \label{eq:dic}
\end{equation}
For a cell-like IGMF, one expects $\delta\theta \propto \epsilon^{-3/4}B\sqrt{\lambda}$ and $\delta t \propto
\epsilon^{-5/2}B^2\lambda$ for $\lambda \ll D_\mathrm{IC}$. 

\subsection{CTA Simulations}
We generate CTA observations by folding the \elmag output spectra 
with the CTA instrumental response function (IRF) for the ``Array~E'' configuration.
The IRF and the expected background rate have been determined from Monte Carlo simulations \citep{bernloehr2013}.
For each source (except 1ES\,0229+200),
 we simulate a $\mathrm{T}_\mathrm{obs} = 20\unit{hours}$
observation under a constant zenith angle of $20^\circ$ and a ratio between source and off-source 
exposure of $\alpha = 0.2$. 
AGN observations of this duration are envisaged 
during the initial years of data taking \citep[][in preparation]{ctakeyscience}. 
Since we assume that 1ES\,0229+200 has the softest intrinsic spectrum
of all considered sources, the observation time is doubled to 40\,hr.

We approximate the CTA PSF as a Heavyside step function 
that is nonzero within its 80\,\% containment radius $r_{80}$. 
Hence, we discard all \elmag output photons with $\delta\theta > r_{80}$
(indicated by the red solid line in the top panel of Figure \ref{fig:2dhist}).
Furthermore, we assume that the sources have been active for 
10\,yr, approximately the time that \gray sources have been observed
(red solid line in the bottom panel of Figure \ref{fig:2dhist}). 
This cut on the delay time effectively supersedes the PSF cut, 
since photons arriving with $\delta\theta > r_{80}$
usually have delay times $\delta t \gg 10$\,yr. 
We have verified this with the \elmag simulations for all considered 
sources and three configurations of the magnetic field discussed in Sec. \ref{sec:results}.
In Section \ref{sec:results}, we also examine the impact on the results if the  cut on the delay time is relaxed.
The remaining fraction of the \elmag output spectra is then interpolated 
with a cubic spline in order to guarantee a smooth spectrum for the CTA simulation.
The spectra are rescaled, so that the integrated flux above a certain energy threshold 
$E_\mathrm{thr} > \epsilon_\mathrm{max}$ is independent 
of $(B,\lambda)$. 
The assumed integrated fluxes  and
values for $E_\mathrm{thr}$ are listed in Table \ref{tab:srcs}.
For PG\,1218+304, the integrated flux 
and $E_\mathrm{thr}$ are chosen to match the VERITAS observation of this source (\citeauthor{1es1218veritas2013},  for the VERITAS Collaboration \citeyear{1es1218veritas2013}).
For 1ES\,0229+200 we chose the values obtained with H.E.S.S. observations \citep{1es0229hess2007}.
For PMN\,J1548--2251 and 1RXS\,023832.6--311658 we assume values 
similar to 1ES\,0229+200, namely, that the integrated flux 
above 500\,GeV is equal to 2\,\% of the integrated 
flux of the Crab Nebula (C.U.),
whereas for B2\,0806+35 we take $F = 5\,\%\unit{C.U.}$ above 1\,TeV
(the Crab nebula spectrum is taken from \citealt{crabhess2006}).
All the assumed values are compatible with the results from the 2FHL
except for B2\,0806+35 and an IGMF close to zero.
In this case, the cascade emission is also
 in mild tension with the flux upper limit obtained from VERITAS observations \citep{veritas2016:agnul}.
 
We further note that the extrapolation of the intrinsic 
spectra up to $E_{\tau = 5}$ yields luminosities $L$ that are 
consistent with the requirement that the radiation power $P$
of the jet is less than half the Eddington
 luminosity $P \sim L / 4\Gamma_\mathrm{L}^2 \leqslant L_\mathrm{edd} / 2$  \citep{bonnoli2011},
where we again approximated $\delta_\mathrm{D} \sim \Gamma_\mathrm{L}$. 
Following \citet{meyer2012}, 
we derive the luminosity by integrating the intrinsic spectra
between 50\,GeV and $E_{\tau = 5}$ and multiply the integral by 
 $(1 - z)^{2-\Gamma}\times4\pi d_L^2$. The first factor
accounts for the $K$-correction and $d_L$ is the luminosity distance. 
Generically assuming black hole masses of $10^{8.5}M_\odot$, 
we find that $P$ is at most $\sim$0.2\,\% of $L_\mathrm{edd}$ for PG\,1218+304.

\begin{figure*}
\centering
\ifpdf
\includegraphics[width = 0.49\linewidth]{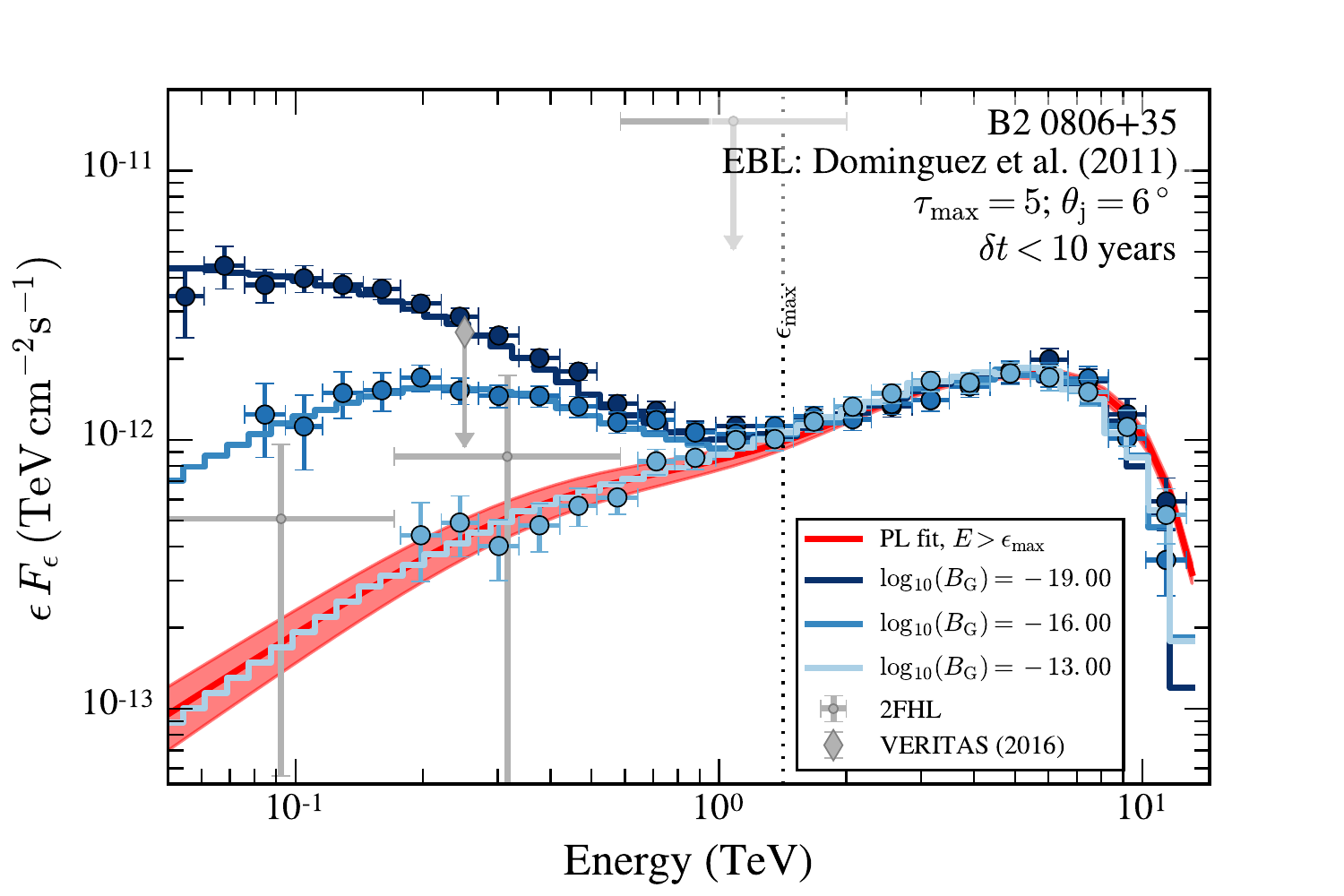}
\includegraphics[width = 0.49\linewidth]{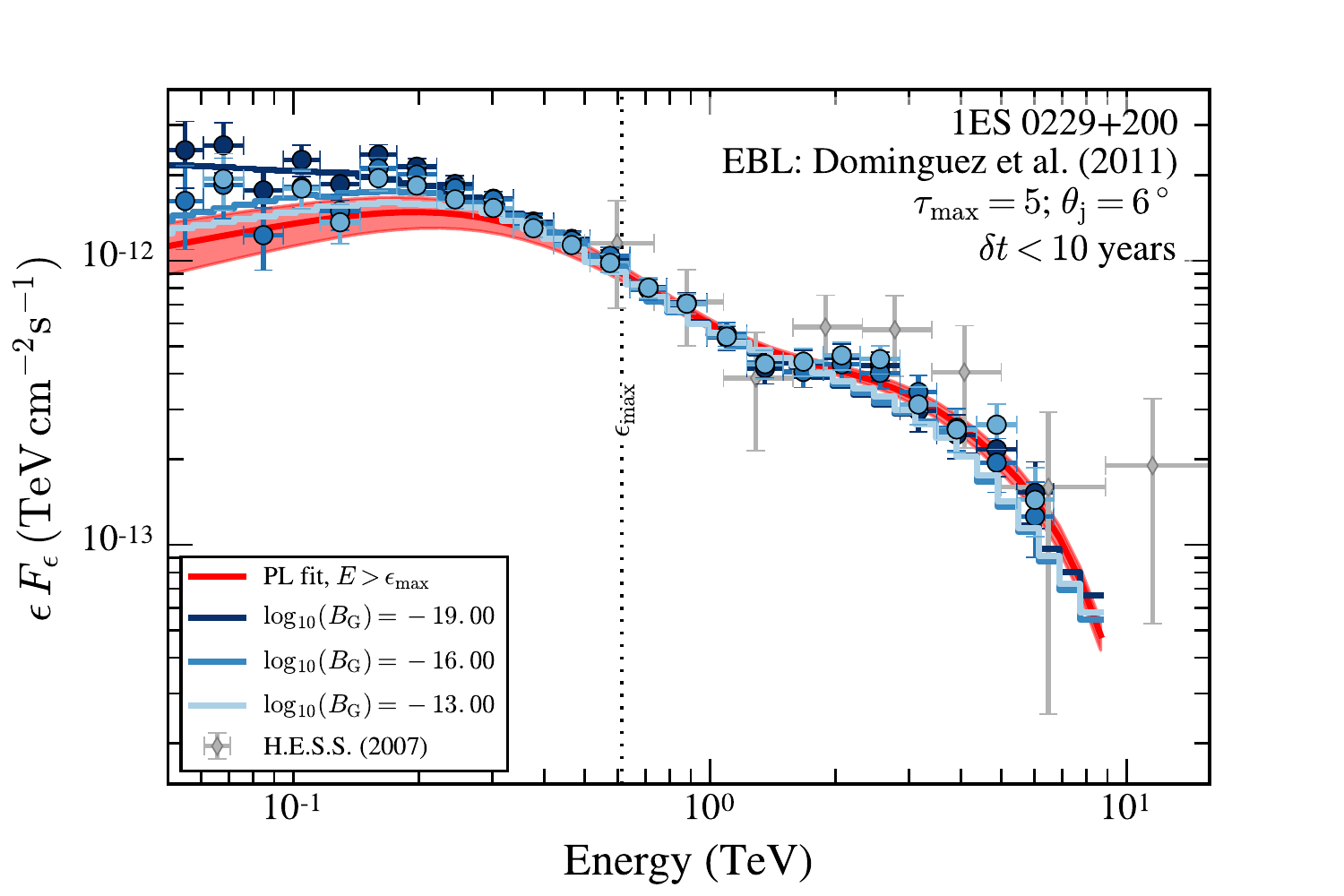}
\includegraphics[width = 0.49\linewidth]{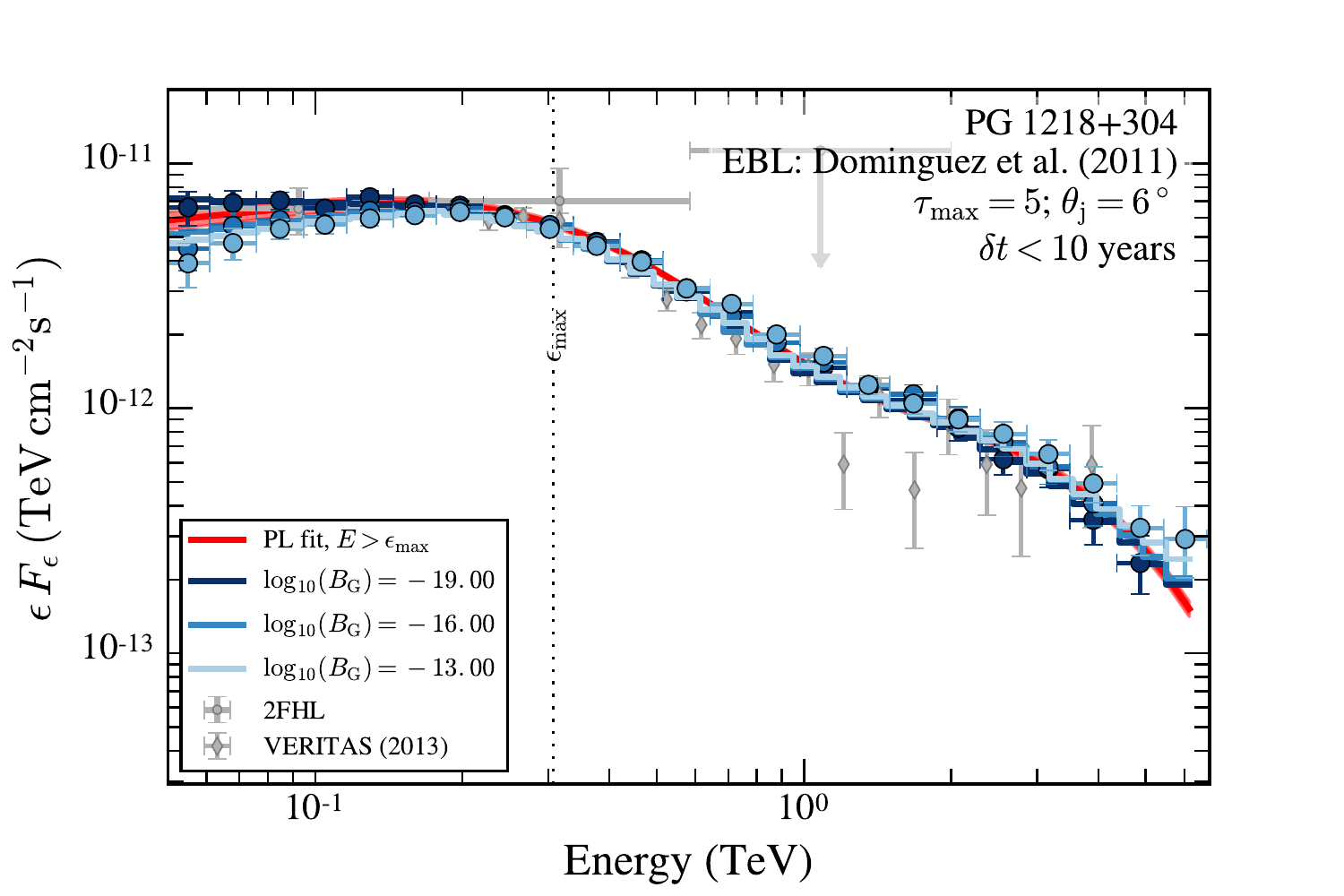}
\includegraphics[width = 0.49\linewidth]{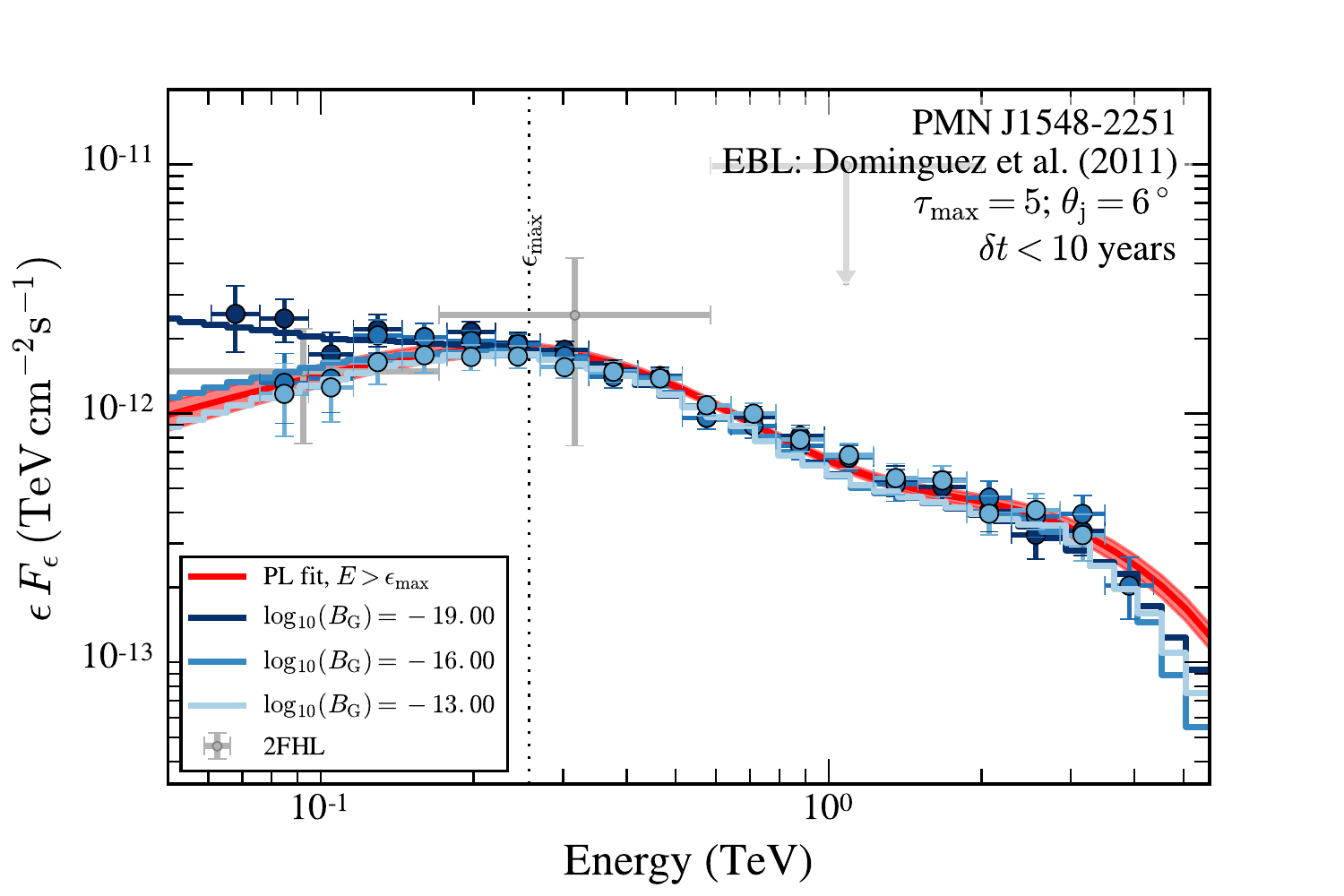}
\includegraphics[width = 0.49\linewidth]{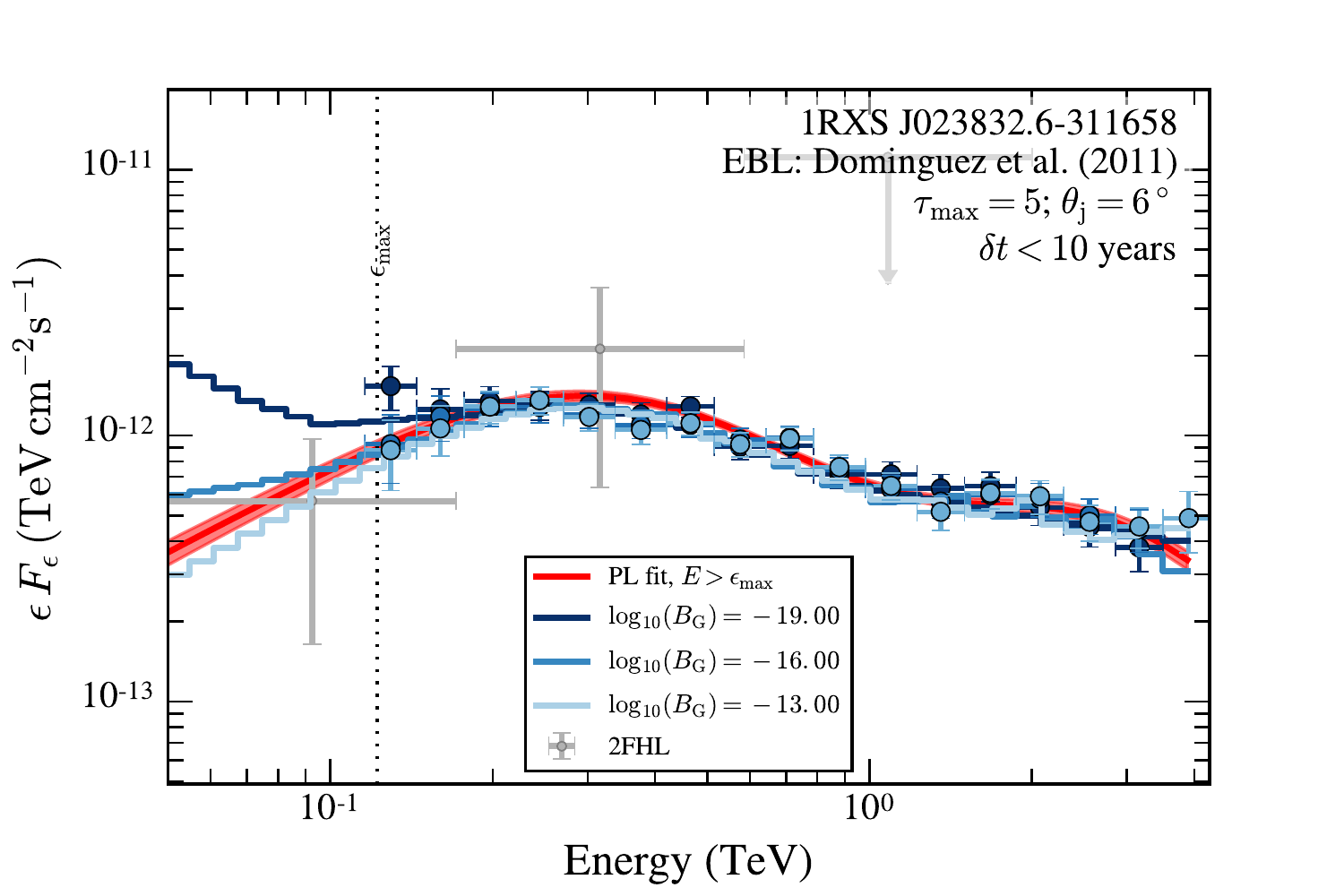}
\else
\vspace{10pt}
\includegraphics[width = 0.49\linewidth]{fig2_B20806.eps}\hspace{5pt}
\includegraphics[width = 0.49\linewidth]{fig2_1ES0229.eps}\vspace{20pt}
\includegraphics[width = 0.49\linewidth]{fig2_PG1218.eps}\hspace{5pt}
\includegraphics[width = 0.49\linewidth]{fig2_PMNJ1548.eps}\vspace{20pt}
\includegraphics[width = 0.49\linewidth]{fig2_1RXSJ023832.eps}
\fi
\caption{\label{fig:spec}Simulated spectra
for the selected sources assuming different 
magnetic field strengths and a constant coherence length of $\lambda_\mathrm{Mpc} = 1$.
 In addition to the simulated CTA data (blue bullets), we show data
points of the 2FHL and IACT observations where available (gray diamonds and squares). 
We apply both the time and angular separation cut 
to the simulated cascade (blue solid lines). 
The maximum cascade energy $\epsilon_\mathrm{max}$ is 
shown as a black dotted vertical line. A $\chi^2$
power-law (PL) fit including EBL absorption to the simulated data with $\log_{10}(B_\mathrm{G}) = -19$ 
and energy bins $E_i > \epsilon_\mathrm{max}$
is shown with red solid lines. 
}
\end{figure*}

Following \citet{meyer2014cta}, 
the rescaled and interpolated spectra are folded with the IRF and multiplied with the observation 
time to yield the number of expected counts for each source in energy bin $i$, $\mu_{i}$. 
The number of background events $b_i$ is obtained by multiplying
 the background rate derived from Monte Carlo simulations by $\mathrm{T}_\mathrm{obs}$.
Adjacent energy bins in which the source is detected with a significance $S_{i} < 2\,\sigma$ are combined
into one bin
(the significance is evaluated with Eq. (17) of \citealt{lima1983}).
If the significance of the combined bin is still below $2\,\sigma$, the bin is discarded.
We show examples of the rescaled \elmag spectra
and the CTA simulation for all sources and three values of $B$ in Figure \ref{fig:spec}.
For most sources, the 2FHL measurements are consistent with the CTA simulations
even if the cascade excess is present.  

\section{Analysis Method}
\label{sec:analysis}
Higher values of the IGMF strength and coherence length 
will lead to a stronger deflection of the $e^+e^-$ pairs
and cause larger time delays and angular separations of the 
cascade photons. 
With our chosen cuts on $\delta t$ and $\delta\theta$
this implies a diminished cascade flux. 
We use a Poisson likelihood ratio test to determine the compatibility of a magnetic field hypothesis characterized through
the expected number of counts $\mu$ for one set of values $(B,\lambda)$ 
with mock data $D$,
generated under the same or a different hypothesis.
For expected signal counts $\mu_i$ and background counts $b_i$ 
in the $i$th energy bin, the likelihood of observing $x_i$ counts from the sky region 
including a source and $y_i$ counts from a background region is
\begin{equation}
\mathcal{L}(\mu_i,b_i; \alpha | x_i,y_i) = 
\mathrm{Pois}(x_i | \mu_i + b_i)\,\mathrm{Pois}(y_i | b_i /\alpha).
\label{eq:like}
\end{equation}
We only consider energy bins for which we expect
a contribution from the cascade, $E_i < \epsilon_\mathrm{max}$, where $E_i$
is the central energy of each bin. 
We further only select bins for which the detection significance of the source is $S_i > 2\,\sigma$.

The number of expected counts depends on the tested $B$-field hypothesis 
and on the intrinsic \gray spectrum. 
The latter can be determined from a power-law fit (including EBL absorption) 
to the energy bins for which the cascade contribution is negligible, i.e. $E_i > \epsilon_\mathrm{max}$.
Example fits are shown in Figure \ref{fig:spec} as red solid lines. The obtained best-fit parameters
are then independent of the IGMF. 
The fit uncertainty can be incorporated  into 
the likelihood by an efficiency term,
$\beta_{i}$. 
The likelihood for this additional nuisance parameter
 can be assumed to follow a Gaussian, so that the total likelihood becomes
\begin{eqnarray}
\mathcal{L}(\mu_i,\theta_i;\alpha,\sigma_i | x_i, y_i) &=& 
  (2\pi\sigma_{i}^2)^{-1/2}\exp(-(1 - \beta_{i})^2 / 2\sigma_{i}^2) \nonumber\\
&{}&\times\,\mathcal{L}(\beta_i\mu_i,b_i; \alpha | x_i,y_i),
\end{eqnarray}
where $\theta_i = (b_i,\beta_i)$ denotes the nuisance parameters and
 $\sigma_{i}$ is the relative theoretical flux uncertainty in the $i$th energy bin  
from the full covariance matrix of the power-law fit (light-red shaded areas in Figure \ref{fig:spec}). 
We  make the simplifying assumption that the best-fit
 intrinsic spectrum is equal to the input spectrum. 
Further systematic uncertainties can be implemented in a similar way. 

Instead of generating many Monte Carlo realizations for the mock data sets, 
we make use of the so-called Asimov data set,
for which $x$ and $y$ are equal to the expected number of counts \citep{cowan2011}.
Denoting the expected number of counts for the IGMF hypothesis under
which the data are generated with  $\mu^D$, the Asimov data set is
  $x_{i} = \mu^D_{i} + b_{i}$ and $y_{i} = b_{i}/\alpha$.
For each source, 
we combine the likelihoods of all considered energy bins 
and find the profile likelihood by maximizing over the nuisance parameters.
The likelihood ratio test (or test statistic, $\mathrm{TS}$) is then
\begin{equation}
\mathrm{TS}=
-2\sum\limits_{\substack{i\\\ E_{i}\, < \,\epsilon_\mathrm{max} \\ S_{i}\, > \,2\,\sigma}}
\ln\left(\frac{
\mathcal L (\mu_{i},\widehat{\theta_i}(\mu_{i}) ;\alpha,\sigma_i | \mu^D_{i} + b_{i}, b_{i} / \alpha)}
{
\mathcal L (\widehat{\mu}_{i},\widehat{\theta}_i ;\alpha, \sigma_i | \mu_{i}^D + b_{i}, b_{i} / \alpha)
}\right).\label{eqn:loglRatio}
\end{equation}
By virtue of the Asimov data set, the maximum likelihood estimators
are simply
$\widehat{\mu}_i = \mu_{i}^D$ and $\widehat{\theta}_i = (b,1)$.
In the numerator, the likelihood is maximized for fixed $\mu_{i}$ in terms of
the background counts and efficiency to yield $\widehat{\theta}(\mu_{i})$ \citep{rolke2005}.

Applying Wilks' theorem, 
the test statistic should asymptotically follow a $\chi^2$ distribution 
with 2 degrees of freedom $\nu$ for the two model parameters $(B,\lambda)$. 
This allows us to convert the $\mathrm{TS}$ values into a significance $p_{\chi^2_\nu=2}$,
with which we can exclude a magnetic field hypothesis for a given mock data set. 
To improve the sensitivity, the likelihoods of the different sources 
are combined by adding the $\mathrm{TS}$ values. 

Since we do not know the IGMF morphology realized in nature, we generate mock data sets for specific scenarios of the 
IGMF strength and coherence length, yielding $\mu^D$ for 
each energy bin and source (see Section \ref{sec:results}). 
The $\mathrm{TS}$ values are then computed 
with respect to the number of expected counts $\mu$ 
for all considered ($B$,$\lambda$) values, which are
extracted from the cascade simulations. 
We thus obtain 
the significances  $p_{\chi^2_\nu=2}$ to rule out
 IGMF morphologies different from the one assumed in each scenario. 

\section{Results}
\label{sec:results}
We generate mock data samples for three different representative 
IGMF configurations: $D_1 = (B_\mathrm{G} = 10^{-15}; \lambda_\mathrm{Mpc} = 10^{-6})$,
$D_2 = (B_\mathrm{G} = 10^{-16}; \lambda_\mathrm{Mpc} = 1)$, and 
$D_3 = (B_\mathrm{G} = 10^{-13}; \lambda_\mathrm{Mpc} = 10)$.
The hypothesis $D_1$ corresponds to the case 
where the IGMF is of primordial origin \citep[e.g.][]{durrer2013} 
and is still allowed by the limits derived by \citet{finke2015}. 
The small values of  $\lambda$
will suppress large deflection of the $e^+e^-$ pairs. 
On the other hand, the values of $D_3$,
close to the IGMF configuration suggested by 
observations of the diffuse \gray background \citep{chen2015helical},
 will lead to large deflections, and most cascade photons will arrive with 
large time delays and outside $r_{80}$. 
The $D_2$ scenario corresponds to an intermediate
case in terms of deflections and tests the hint for a nonzero IGMF 
deduced from evidence of  
 pair halos in \fermiLAT data \citep{chen2015ph}. 
Fields with such values of $\lambda$ could be generated by outflows 
from AGNs \citep[e.g.][]{furlanetto2001}.

\begin{figure*}
\centering
\ifpdf
\includegraphics[width = .9\linewidth]{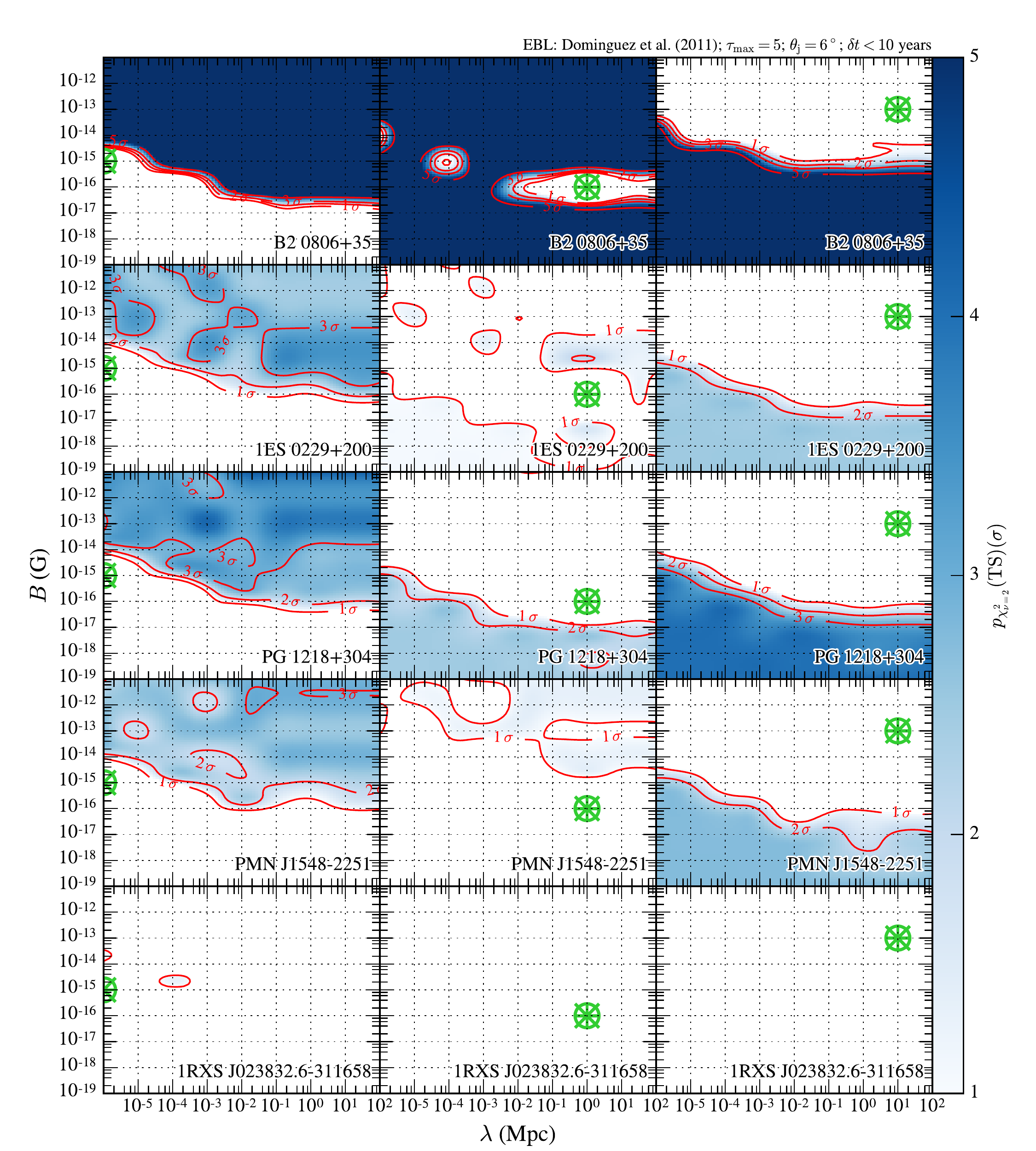}
\else
\vspace{10pt}
\includegraphics[width = .9\linewidth]{fig3.eps}
\fi
\caption{\label{fig:single_sources}Exclusion regions for single-source observations (top to bottom) and 
different IGMF configurations (green markers). 
From left to right, the configurations are $\log_{10}(B_\mathrm{G}) = -15, -16, -13$ with coherence lengths
$\log_{10}(\lambda_\mathrm{Mpc}) = -6, 0, 1$. 
The results on the parameter grid are interpolated with a bivariate spline.
} 
\end{figure*}

Figure \ref{fig:single_sources} shows the possible limits 
in the $(B,\lambda)$ plane obtained 
from the observation of each source alone for $D_j$, $j=1,2,3$ and 
the fiducial set of model parameters.
Both cuts on the angular separation and delay time are applied.
The most constraining limits come from the 
simulated observation of B2\,0806+35. 
The assumptions for the intrinsic spectrum are 
the most optimistic of our source sample as we extrapolate 
the 2FHL spectrum up to  $E_{\tau = 5} \sim 14$\,TeV, albeit the smallest
ratio of X-ray to radio flux (compare Table \ref{tab:srcs}). 
This high-energy cutoff causes a plenitude of cascade photons in the energy range of CTA. 
Such an observation would lead to strong bounds on the IGMF, with possible 
exclusions beyond the $5\,\sigma$ confidence level.
On the other hand, for the highest-redshift source, 
and therefore smallest maximum energy $E_{\tau = 5} \sim 4$\,TeV,
no IGMF value can be ruled out. This is already obvious from Figure~\ref{fig:spec}:
the source is too faint below $\epsilon_\mathrm{max}$
to distinguish between the different IGMF scenarios. 
In the case of 1ES\,0229+200 the fit uncertainties 
are large, and therefore no exclusions are possible if $\mathrm{T}_\mathrm{obs} = 20\,$hours.
Doubling the observation time leads to mild exclusions, as visible in Figure~\ref{fig:single_sources}.
The two remaining sources give similar constraints at the $2\,\sigma-3\,\sigma$ level for $D_1$ and $D_3$. 
For $D_2$, the differences in the spectra are not pronounced enough
to rule out IGMF configurations leading to either smaller or larger deflections.
As expected, the limits are independent of the 
coherence length as long as $\lambda \gg D_\mathrm{IC}$. 
The small structures in the exclusion plots can be explained with the intrinsic scatter of the \elmag
simulations. 

\begin{figure*}
\centering
\ifpdf
\includegraphics[width = 1\linewidth]{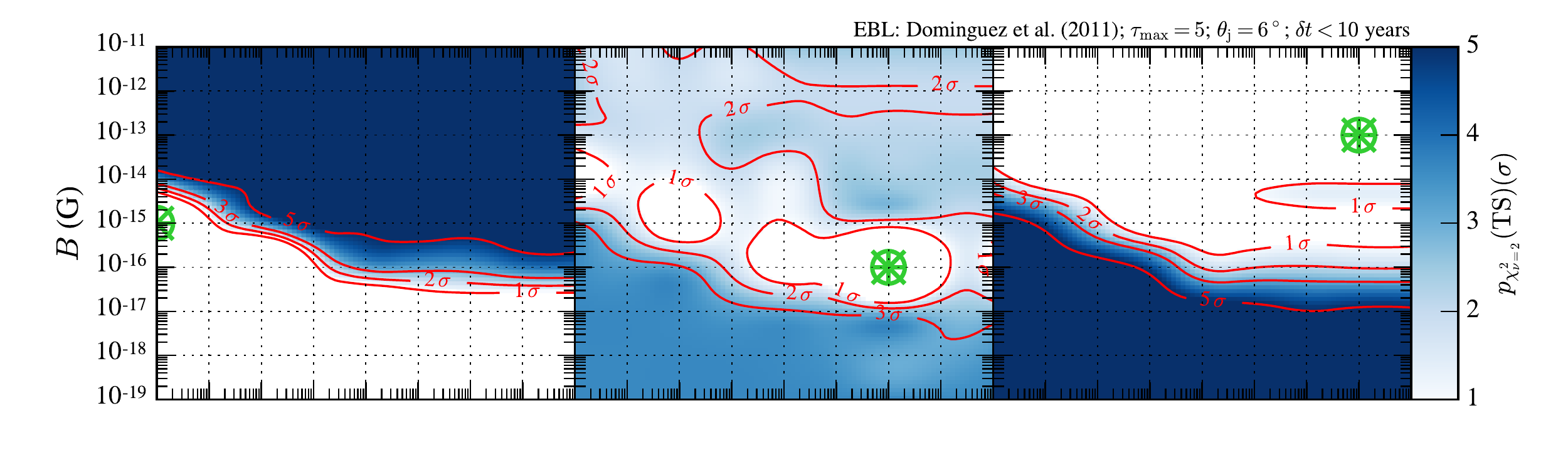}\vspace{-20pt}
\includegraphics[width = 1\linewidth]{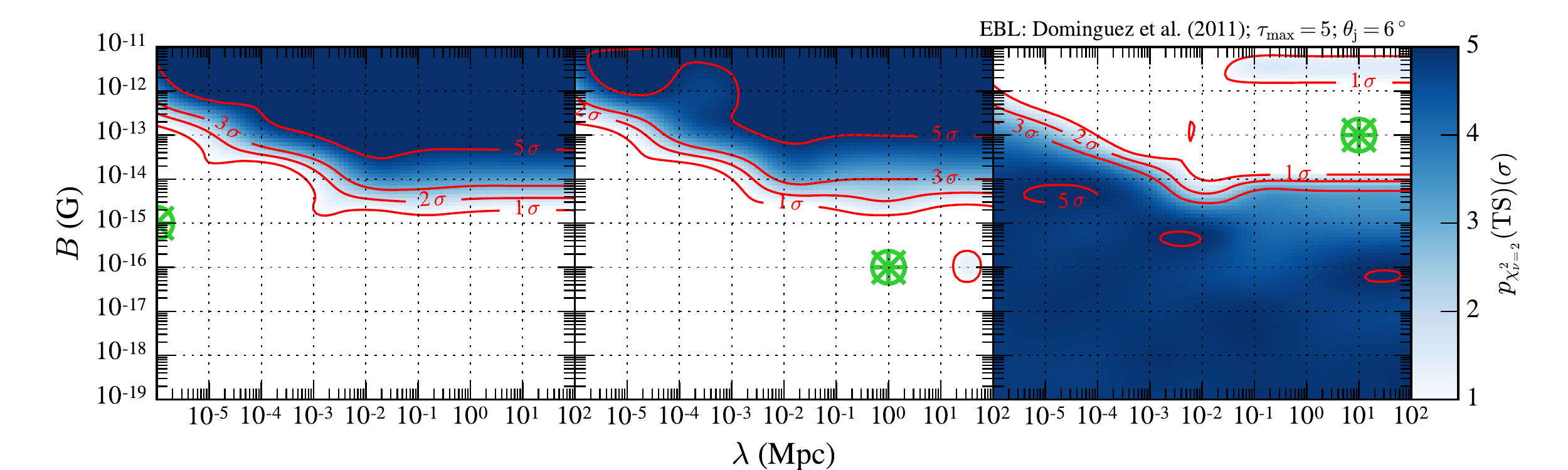}
\else
\vspace{10pt}
\includegraphics[width = 1\linewidth]{fig4_fid.eps}\vspace{10pt}
\includegraphics[width = 1\linewidth]{fig4_time.eps}
\fi
\caption{\label{fig:combined}Exclusion plots 
from a combined likelihood of all blazars except B2\,0806+35 for three tested IGMF configurations.
Top: limits with a maximum time delay $\delta t < 10$\,years.
Bottom: limits without any restrictions on the delay of the cascade photons.
}
\end{figure*}

We combine the likelihoods of the sources and reevaluate the 
exclusion regions (Figure \ref{fig:combined}, top panels). 
We exclude the source B2\,0806+35, 
as we consider its cascade yield uncertain due to 
the extrapolation of the intrinsic spectrum 
by more than one order of magnitude and its small 
 $F_\mathrm{X} / F_\mathrm{R}$ ratio.
The limits considerably improve for all tested hypotheses.
If the IGMF configurations close to $D_3$ are realized in nature, 
CTA observations could improve current limits \citep{finke2015} by two orders
of magnitude.  
The $D_1$ case results in stringent upper bounds on the  
IGMF, and field strengths with  
 $B_\mathrm{G}\gtrsim10^{-14}$
 can be excluded at high significance, independent of the coherence length.
The reason is that most cascade photons still arrive within 
the containment radius and with small time delays, leading to 
a large excess at GeV energies. For higher IGMF strengths
or coherence lengths, the excess decreases, which is incompatible with this particular mock data set.
Therefore, upper limits on $B$ are obtained. 
This situation will always occur for IGMF scenarios 
that lead to minimal deflections of the cascade photons. 
For $D_2$,  the combined likelihood leads to an 
exclusion of magnetic fields smaller
than $10^{-17}\unit{G}$ at $3\,\sigma$. 
Larger fields are at tension with the data at $1\,\sigma-2\,\sigma$. 
As for the single-source limits, the combined limits 
only show a dependence on the coherence length if $\lambda_\mathrm{Mpc} 
\lesssim 10^{-2}$.


Relaxing the cut on the maximum delay time 
has a strong impact on the sensitivity to the detection of cascade photons
(Figure \ref{fig:combined}, bottom panels).
Applying no cut on the maximum time delay at all is certainly an oversimplification 
due to finite AGN lifetimes, which are 
estimated to lie between $10^6$ and $10^8$ yr \citep[e.g.,][]{parma2002}. 
However, we do not expect a significant change if we would instead 
assume $\delta t < 10^8$\,yr, 
as even for the tested IGMF leading to the largest delays $(B_\mathrm{G} = 10^{-11}, \lambda_\mathrm{Mpc} = 100)$, 
a significant fraction of the cascade photons still arrive 
with smaller delays. 
Interestingly, comparing the $\delta t < 10$\,yr case to the case with no time cut, 
one sees that for configurations leading to small deflections (as in $D_1$), 
the projected limits worsen by two orders of magnitude. 
The reason is that more cascade photons 
reach the observer since the $\delta t$ cut is more stringent
than the requirement $\delta\theta < r_{80}$ (see Figure \ref{fig:2dhist}).
As a result, increasing $B$ or $\lambda$ will have 
a weaker effect on the spectra up to the point where $\delta\theta > r_{80}$.
Consequently, configurations with $B_\mathrm{G} \lesssim 10^{-15}$
cannot be distinguished with high significance from a zero IGMF regardless of $\lambda$.
This also explains the differences in $D_2$ and $D_3$.
Only in the $D_3$ scenario does the applied cut on $\delta t$ 
lead to conservative limits (right panels of Figure \ref{fig:combined}).
In this case, the time cut removes a large number
of cascade photons so that the cascade bump is less pronounced 
even for small values of the magnetic field.

\begin{figure*}
\centering
\ifpdf
\includegraphics[width = 1\linewidth]{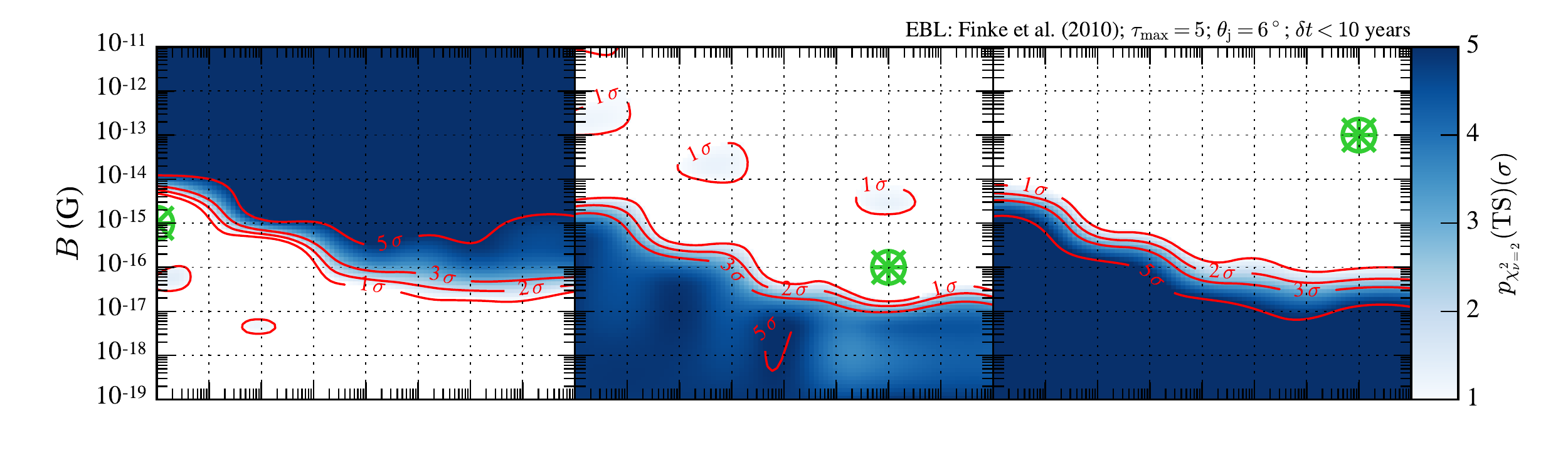}\vspace{-20pt}
\includegraphics[width = 1\linewidth]{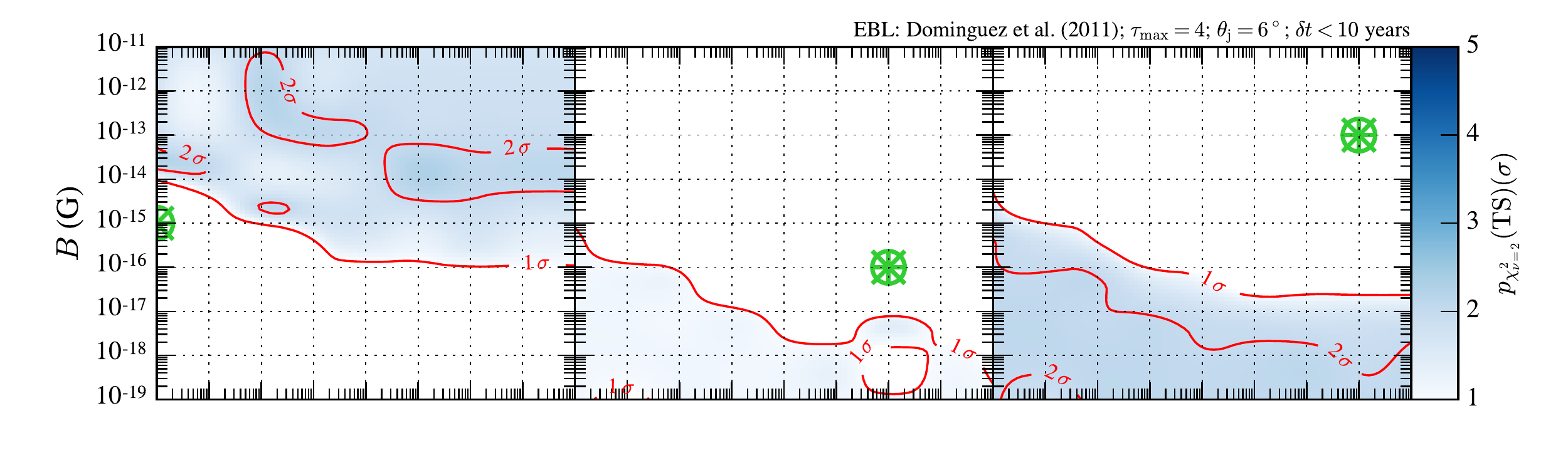}\vspace{-20pt}
\includegraphics[width = 1\linewidth]{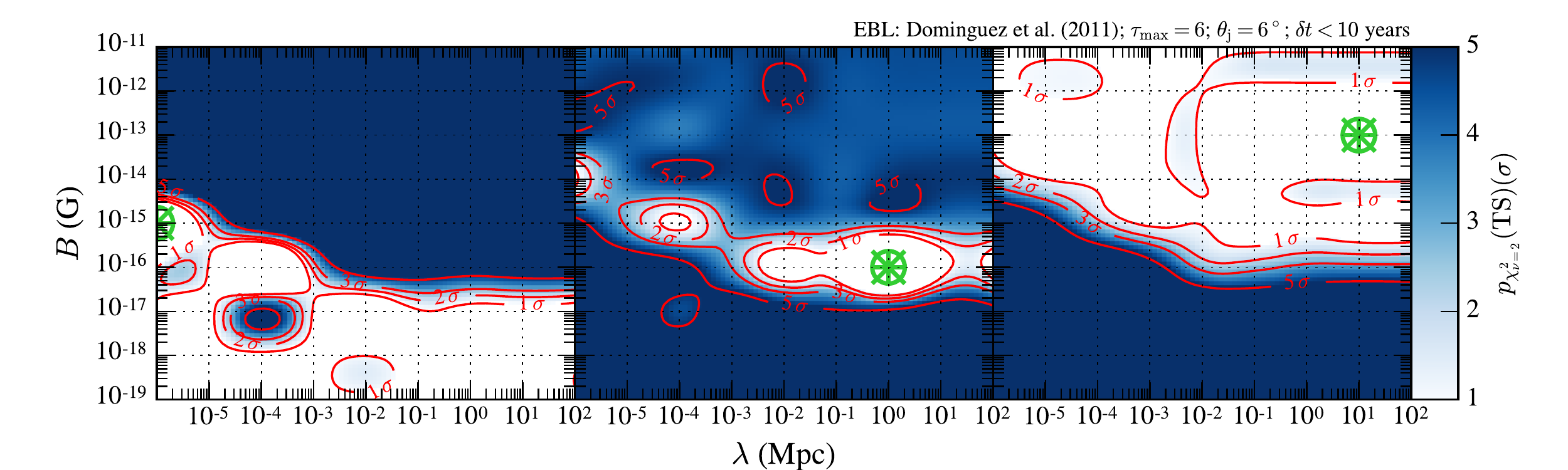}
\else
\vspace{10pt}
\includegraphics[width = 1\linewidth]{fig5_finke.eps}\vspace{10pt}
\includegraphics[width = 1\linewidth]{fig5_tau4.eps}\vspace{10pt}
\includegraphics[width = 1\linewidth]{fig5_tau6.eps}
\fi
\caption{\label{fig:sys}
Same as Figure \ref{fig:combined} with $\delta t < 10$\,yr
but for the \citet{finke2010} EBL model and $\tau_\mathrm{max} = 5$ (top),
and for the \citet{dominguez2011} EBL model with $\tau_\mathrm{max} = 4$ (middle)
 and $\tau_\mathrm{max} = 6$ (bottom).
}
\end{figure*}

We further investigate the dependence 
of the projected limits on the chosen EBL model 
and energy of the spectral cutoff in Figure \ref{fig:sys}.
In the top panels, the EBL model 
of \citet{finke2010} is used instead of the 
photon density predicted by \citet{dominguez2011}, while $E_{\tau = 5}$ is held constant. 
For the \citeauthor{finke2010} model, 
the limits are strengthened since the attenuation 
is slightly higher for sources with $z \gtrsim 0.1$.
As the attenuation increases, so does the number distribution of the $e^+e^-$ pairs, which
is given in the steady-state approximation by 
\begin{equation}
N(\gamma) = |\dot{\gamma}|^{-1} \int_\gamma^\infty d\gamma'Q(\gamma'),\label{eq:eldistr}
\end{equation}
where $\dot{\gamma} = c\gamma^3 / D_\mathrm{IC} \propto \gamma^2$ is the energy loss of the pairs due to 
IC scattering and $Q(\gamma) = \mathrm{d}N / \mathrm{d} E (1 - \exp(-\tau))$
is the injection rate with $E = 2m_ec^2\gamma$.
For hard intrinsic \gray spectra and since $\tau$ increases rapidly with energy, the integral of Eq. \eqref{eq:eldistr} 
will be almost independent of the lower integration bound and
$N(\gamma)$ is dominated by $|\dot{\gamma}|^{-1}$, so that
$N(\gamma) \propto \gamma^{-2}$.
Thus, $N(\gamma)$ is dominated by low energy pairs \citep{tavecchio2011}.
For $\gamma = 5\times10^4$ (corresponding to a 50\,GeV \gray), $N(\gamma)$
is about 4\,\%--5\,\% larger for the \citet{finke2010} model. 

Considering instead the 
EBL model of \citet{dominguez2011} but lowering
the maximum spectral energy to the value that corresponds to $\tau_\mathrm{max} = 4$ 
decreases the sensitivity significantly
(middle panels of Figure \ref{fig:sys}).
The situation is reversed if we increase the maximum energy 
so that $\tau_\mathrm{max} = 6$ (bottom panels of Figure \ref{fig:combined}).\footnote{
The corresponding energies are 10.6, 8.0, 7.5, and 5.6\,TeV
for 1ES\,0229+200, PG\,1218+304, PMN\,J1548--2251, and 1RXS\,J023832.6--311658,
respectively.} 
In this case, the IGMF strength could be determined within one order of
magnitude in the $D_2$ scenario. 
These findings underline the necessity that the intrinsic spectra 
need to extend to energies as high as possible in order to derive strong 
constraints on the IGMF. 

We have also tested the dependence on the jet opening 
angle, and even a highly collimated jet with $\theta_\mathrm{j} = 1^\circ$
has a negligible effect on the limits.

Throughout this work, we have assumed 
that the intrinsic blazar spectrum is given by 
a simple power law.
In principle, the cascade component could be mimicked by
features in the intrinsic blazar spectrum caused by, e.g., 
multiple \gray-emitting regions \citep[e.g.][]{lefa2011:blob}.
However, the multiwavelength emission
of EHBLs can also be successfully described with one emission zone 
under the assumption of very high minimal Lorentz factors of the 
underlying electron distributions or 
 electron distributions with a Maxwellian shape 
\citep{katarzynski2006,tavecchio2009,kaufmann2011,lefa2011:hard,
bonnoli2015}.  
Furthermore, the sensitivity for the cascade 
has been derived here from a combined likelihood of several sources.
It would appear highly contrived if the intrinsic 
spectra of multiple sources showed features at exactly the
right energies where one expects the cascade component 
for a given IGMF. 

\section{Conclusions}
\label{sec:conclusion}
Utilizing a standard likelihood 
ratio test,  
future observations with CTA of a small number of certain HBLs 
will yield strong constraints 
of the IGMF. 
Especially HBLs with hard spectra that extend 
to energies corresponding to an optical depth of $\tau\gtrsim5$ are particularly well suited since a large amount of energy will be reprocessed in the electromagnetic cascade.
The large energy range covered with CTA makes it possible 
to probe the EBL cutoff \citep{mazin2013} 
and to ensure the absence of an intrinsic spectral break. 
Simultaneously, one can search for cascade photons at the low-energy end of the spectra.
 Throughout this article, we have assumed the
``Array E'' configuration of CTA \citep{bernloehr2013}. 
The final constraints derived from real data will
depend on the actually realized configuration 
and data analysis.  

Nevertheless, the (non)observation of a cascade excess in CTA 
spectra will allow us to limit an IGMF with a high (low) field strength 
depending on the actual IGMF morphology realized in nature, the 
maximum emitted \gray energies, and the duty cycles of the considered
sources.
CTA observations should be able to either confirm or rule out evidence of a nonzero IGMF
 \citep[e.g.][]{essey2011,chen2015ph,chen2015helical}
and improve current limits on its field strength and coherence length 
by orders of magnitude \citep[e.g.][]{taylor2011,arlen2014,finke2015}.
Especially if IGMF configurations are realized in nature that lead to either 
strong or very small deflections of the $e^+e^-$ pairs,
it will be possible to rule out large fractions of the IGMF parameter space. 
Magnetic fields of the order of $B \sim 10^{-16}\unit{G}$
with coherence lengths $\lambda \sim 1\unit{Mpc}$ will 
be more difficult to constrain as the 
cascade radiation will only lead to a slight excess over the primary \gray emission.

CTA observations will in general not be able 
to distinguish between a primordial and astrophysical 
origin of the IGMF. 
Furthermore, the sensitivity estimates 
depend strongly on the assumed 
 cutoff energy of the spectra.
Spectra extending only up to energies
so that $\tau_\mathrm{max} = 4$ 
will not generate sufficient cascade radiation to constrain the IGMF.  
If, on the other hand,  the spectra reach very high energies with optical depths $\tau \sim 6$,
primordial IGMF scenarios could be ruled out given that
 the coherence length is $\lambda \gtrsim 0.1\,$Mpc.
The projected limits also strongly depend on the assumed 
\gray activity time of the AGN. 
The maximum allowed delay time of cascade photons is degenerate with the 
IGMF strength, and small values of $\delta t$ only yield conservative 
limits if small values of values of the IGMF strength are to be constrained. 
It should be noted that the cell-like morphology of the IGMF adopted in the \elmag code
neglects the dependence of the limits on the actual 
IGMF power spectrum \citep{caprini2015}. 
Especially for red power spectra, 
the cell-like assumption breaks down and 
lower limits on the IGMF have to be relaxed. 

We have only used photons arriving 
within the 80\,\% containment radius of the PSF. 
In future work, the analysis should be extended 
to incorporate the extended pair-halo emission. 
This will add further information to the likelihood 
and will make it easier to distinguish between IGMF
scenarios.
Interestingly, due to the time delay of the cascade photons, 
such halos could still be present even if the source
already ceased its activity \citep{neronov2010ph,inoue2011}
and could be searched for the 
envisaged CTA extragalactic survey.

\section*{Acknowledgements}
This paper has gone through internal review by the CTA Consortium.
The authors would like to thank Anthony Brown, 
Jonathan Biteau, Matteo Cerruti, Michele Doro, Susumu Inoue, 
Kohta Murase, Elisabete de Gouveia Dal Pino,
Vitor de Souza, and especially Alberto Dom\'inguez 
for discussions and comments on the manuscript.  
J.C. is a Wallenberg Academy Fellow.

\software{ELMAG \citep{elmag}}
\bibliography{IGMF,vhe_spectra}

\end{document}